# Plasmonic nanoantennas: enhancing light-matter interactions at the nanoscale

Shobhit K. Patel, and Christos Argyropoulos*

Department of Electrical and Computer Engineering, University of Nebraska-Lincoln, Lincoln, NE, 68588, USA

*christos.argyropoulos@unl.edu

**Abstract:** The research area of plasmonics promises devices with ultrasmall footprint operating at ultrafast speeds and with lower energy consumption compared to conventional electronics. These devices will operate with light and bridge the gap between microscale dielectric photonic systems and nanoscale electronics. Recent research advancements in nanotechnology and optics have led to the creation of a plethora of new plasmonic designs. Among the most promising are nanoscale antennas operating at optical frequencies, called nanoantennas. Plasmonic nanoantennas can provide enhanced and controllable light-matter interactions and strong coupling between far-field radiation and localized sources at the nanoscale. After a brief introduction of several plasmonic nanoantenna designs and their well-established radio-frequency antenna counterparts, we review several linear and nonlinear applications of different nanoantenna configurations. In particular, the possibility to tune the scattering response of linear nanoantennas and create robust optical wireless links is presented. In addition, the nonlinear and photodynamic responses of different linear and nonlinear nanoantenna systems are reported. Several future optical devices are envisioned based on these plasmonic nanoantenna configurations, such as low-power nanoswitches, compact ultrafast light sources, nanosensors and efficient energy harvesting systems.

## 1 Introduction

Antennas are usually referred to transducers which can convert electrical power into propagating electromagnetic waves and vice versa [1, 2]. Owing to the fundamental importance of this functionality, their use at microwave and radio frequencies (RF) is applied in several devices around us. Their design has seen major developments starting from the invention of wireless telegraphy by Marconi around the turn of 19th century. During the past decades, the rapid development of electronics and wireless communications led to greater demand for wireless devices that can operate at different frequency bands, such as mobile telecommunications systems, Bluetooth devices, wireless local area networks, and satellite communications. Additionally, several communication devices, e.g. cell phones, demand the antenna apparatus to be very compact and confined in an ultrasmall dimensional footprint [3-11]. These two requirements have triggered extensive research on the design of compact and complex antenna designs for RF and microwave frequency applications.

On a different context, recent developments in nanoscale optical systems have generated a major interest in the optical counterpart of the well-established RF/microwave antennas, the so-called plasmonic nanoantennas. The design of these metallic nanoantennas is different from the well-established RF/microwave antennas in two important aspects. First, the assumption of metals behaving as perfect electrical conductors (PEC), which is usually followed in microwave frequencies, is no longer valid at optical frequencies. The electric fields can penetrate inside metals at this high frequency regime and the Ohmic losses are substantially higher compared to microwave frequencies. Second, metallic and dielectric interfaces can sustain surface plasmon polaritons (SPP) waves in the visible confined at nanoscale regions [12, 13]. Therefore, the response of the aforementioned optical plasmonic nanoantenna structures is drastically different compared to their microwave counterparts. The most notable example among these differences is the strong field confinement obtained at the subwavelength nanogap region of metallic nanoantennas. This nanogap region can be used as feeding or receiving point of the nanoantenna system [14]. Additionally, new physical phenomena arise at nanoscale light-matter interaction regions. Their theoretical study requires the development of new analytical and modeling tools to account for deviations from the classical RF antenna theory to the optical and quantum response of plasmonic nanoantennas. As a result, serious efforts have been recently devoted to extend the well-established and thoroughly studied concepts of radio frequency antennas to their optical counterparts [15-22].

## 2 Comparison between RF and optical antennas

Figures 1a-1b, 2a-2b, and 3a-3b demonstrate a schematic comparison between radio frequency antennas [7, 23] and their optical nanoscale counterparts [24, 25]. In general, RF antennas have dimensions on the order of several centimeters. They are extremely important devices in modern wireless communications. On the other hand, optical nanoantennas, with typical dimensions on the order of a few hundred nanometers, have several important optical applications due to their ability to achieve extremely high field values localized in subwavelength "hotspots".

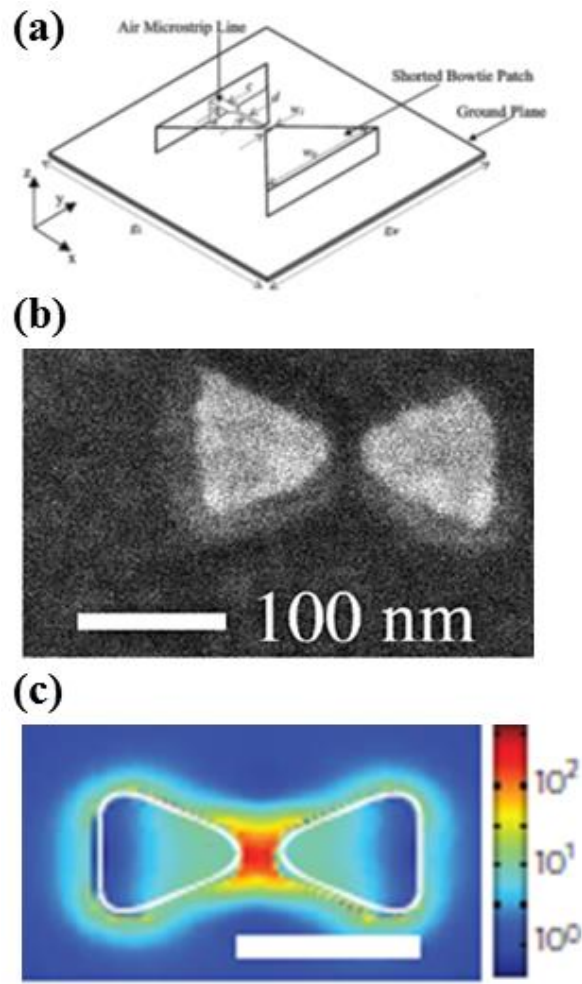

**Figure 1.** Examples of bowtie (a) RF planar antenna [23] and (b) optical nanoantenna [24]. (c) The intensity enhancement of the bowtie nanoantenna shown in (b) [25]. The scale bar in this caption is 100nm.

An example of a broadband planar RF bowtie antenna is shown in Figure 1a [23]. Drawing inspiration from its radio-frequency counterpart, the optical bowtie nanoantenna is presented in Figure 1b. It is made of two gold nanotriangles separated by an ultrasmall nanogap (30nm) [26]. Figure 1c demonstrates the field intensity enhancement localized in the gap region of this bowtie nanoantenna. The field enhancement was calculated with full-wave electromagnetic simulations [24]. Very strong field intensity is obtained which is localized in the nanogap region of the nanoantenna. Metallic bowtie nanoantennas can be used to significantly improve the coupling mismatch between external light radiation and nanoscale emitters or receivers placed at the nanogap [26]. However, the fabrication and characterization of these optical nanoantennas is challenging because of their ultrasmall nanoscale dimensions. Furthermore, the high Ohmic losses of metals at optical frequencies pose another severe limitation in their realistic applications. It is interesting that RF bowtie planar antennas have the advantage of broad bandwidth compared to other well-established RF antennas, such as dipoles [2]. On the other hand, bowtie nanoantennas are more narrowband compared to nanodipoles at optical frequencies [24, 26].

Yagi Uda RF and optical antenna designs are presented in Figure 2 [25, 27]. In general, Yagi Uda RF antennas are built by multiple antenna elements out of which one feed element is connected to the transmitter source or receiver circuit [2]. They can achieve strong directivity and are ideal candidates for long-range communication systems. The phase shift of their antenna elements is adjusted so that constructive interference exists only in the forward direction and not in any other direction. They are usually terminated by one reflector element, which is longer than the emitted radiation's wavelength λ, in order to achieve inductive operation at the back side of their geometry. The phase of the current traveling along the reflector lags the voltage induced in it and, as a result, its impedance is inductive. The reflector is always placed at the back side to reduce the backward and enhance the forward direction radiation. The other multiple antenna elements act as radiation directors. They have shorter dimensions than the emitted wavelength λ. These directors operate as capacitors in order to improve the radiation focus in the forward direction. The phase of the current along each director leads the phase of the voltage induced in it. This type of phase distribution across the array of directors leads to constructive interference and ultimately the phase progression of the radiation. The distance between each director element is 0.3λ. In addition, the distance between the feed and reflector is 0.25λ to improve the directivity [2, 27]. This complex antenna configuration can provide substantial increase in directivity compared to a simple dipole.

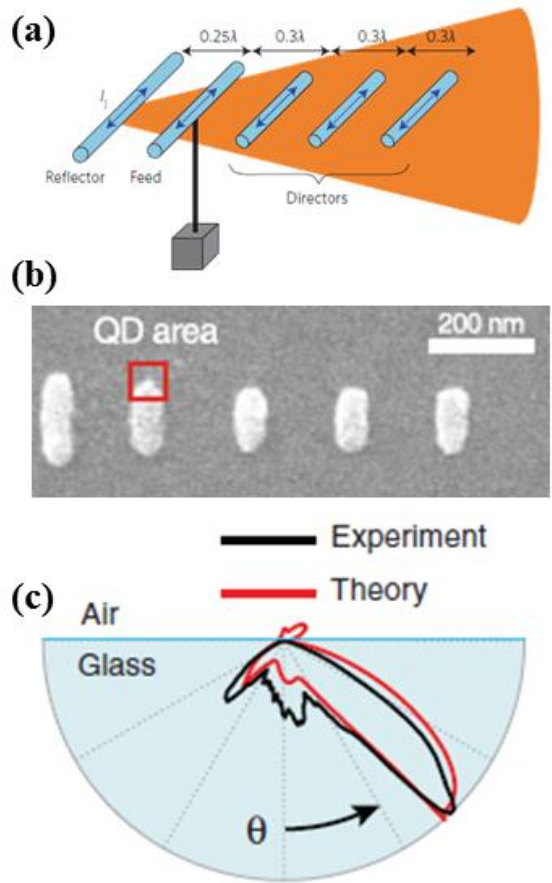

**Figure 2.** Example of Yagi Uda (a) RF antenna [27] and (b) optical nanoantenna [25]. (c) Angular radiation pattern measured (black) and theoretical (red) of the Yagi Uda nanoantenna shown in (b) [25].

To realize a Yagi Uda nanoantenna operating at optical frequencies, five gold nanorods are employed as its elements and are placed over a glass substrate (Figure 2b) [25]. Three of them are used as directors, one is the feed element and one is the reflector. The feed element is driven by a quantum dot emitter [25], which ordinary has a pure dipolar radiation pattern. The plasmonic reflector is inductively detuned to reduce the backward optical radiation. The directors are capacitively detuned to direct the optical wave in the forward direction [25]. The positions of these elements are kept in such a way that the traveling wave is always pointed towards the directors. The angular radiation pattern of the Yagi Uda nanoantenna is presented in Figure 2c. It can be clearly seen that it is a very directional optical nanoantenna [25].

Next, examples of an RF microstrip patch antenna and an optical patch nanoantenna are presented in Figure 3. Specifically, the geometry of the microstrip patch antenna is shown in Figure 3a and its radiation pattern measured at 1.41GHz is presented in Figure 3c [7]. It is composed of a 56.4 ×56.4mm$^2$ square metallic patch placed over a 1.5mm dielectric substrate. The substrate is terminated by a 1mm thick ground plane. Microstrip patch antennas have the advantages of small size, ease of fabrication and low cost, which make them ideal candidates to build compact printed antennas at radio frequencies [3]. They also have few disadvantages, such as moderate gain and narrow bandwidth. Their disadvantages can be overcome by incorporating in their geometry metallic meander elements, inductive slots and metamaterials [7]. The bandwidth of microstrip patch antennas can be increased by mounting the metallic patch on a ground plane with a thicker dielectric spacer [3-5].

In visible frequencies, the optical counterpart of the RF patch antenna is the nanopatch antenna, which consists of a silver nanocube placed over a gold film separated by a thin dielectric layer [28-30]. The geometry of this nanoantenna configuration is presented in Figure 3b and its directional radiation pattern is shown in Figure 3d. The nanocube's side length ranges between 65-95nm and the gap size can range from 5-15nm in this design, as it is presented in Figure 3b. At the resonant frequency, a localized plasmon mode is formed inside the nanogap of this nanoantenna [28-30] and the dominant electric field component in this region is oriented towards the normal z direction (Figure 3b). This particular nanopatch antenna can have field enhancement with maximum values approaching 100 for 5nm gap and 80nm nanocube side length. The resonance of this nanoantenna can be easily tuned from 500nm to 900nm by varying the nanocube size or the nanogap thickness [29, 30]. Note that the nanopatch antenna is less directional, as it is demonstrated in Figure 3d, compared to the multi element Yagi Uda nanoantenna presented in Figure 2c, in a similar way to their RF antenna counterparts.

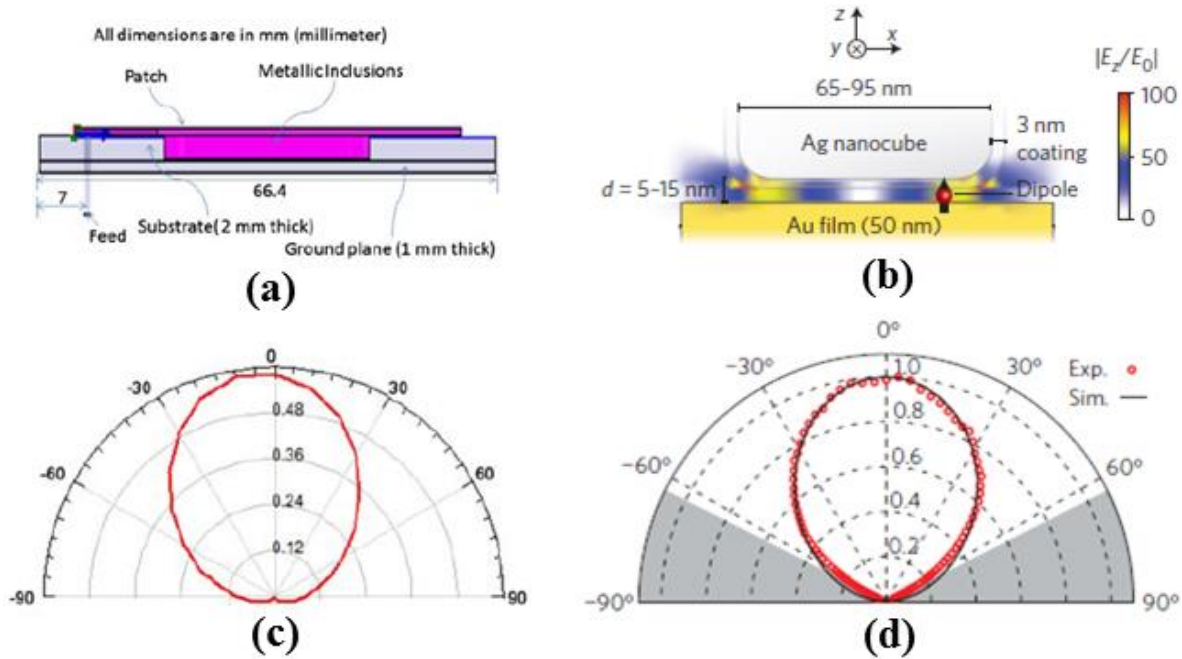

**Figure 3.** (a) RF microstrip patch antenna [7]. (b) Nanopatch antenna built by a gold film coupled to a silver nanocube. The maximum field enhancement is confined at the nanogap of this configuration [28]. (c) Radiation pattern of the RF microstrip patch antenna shown in (a) [7]. (d) Simulated (black) and measured (red) radiation patterns of the optical nanopatch antenna shown in (b) [28].

## 3 Optical plasmonic nanoantennas

Different scientific communities will benefit from the field of optical nanoantennas, as it holds the potential for unprecedented subwavelength light matter interactions, strong coupling between far-field radiation and localized sources at the nanoscale, as well as the exciting possibilities of realizing efficient wireless links between optical nanocircuit components. In the following, we will review several linear, nonlinear and photodynamic optical plasmonic nanoantenna configurations. We will also present tunable and reconfigurable nanoantenna designs.

### 3.1 Linear nanoantennas

Usually, RF and optical antennas have linear, reciprocal and passive response. RF dipole antennas are the simplest and most widely used type among the linear antenna configurations.

Figure 4a depicts a thin cylindrical RF dipole antenna with height h, cylindrical cable radius r (r << h) and gap thickness g. Its cables thickness is much smaller compared to the received or transmitted radiation wavelength (λ>>diameter of conductive cables). The height of this antenna is given by the relation h≈λ/2, where λ is the free space wavelength.

The optical counterpart of the well-established RF dipole antenna is shown in Figure 4b. It is composed of two plasmonic metallic nanorods separated by a nanogap, where inductive or capacitive nanoloads can be placed to achieve tunable performance and impedance matching to the incident radiation. Note that theoretical and experimental studies of different optical nanodipole configurations have been presented by many research groups during the recent years [17, 31-42]. The dimensions of the current nanodipole antenna are: h = 110nm, r = 5nm and g = 3nm. Interestingly, the height h of the nanodipole antenna at its resonant frequency 356THz, which corresponds to an approximate wavelength λ=840nm, is considerably shorter than one-half of the incident light's wavelength [15]. This is in contradiction with classical RF antenna theory. In traditional dipole antenna designs, the antenna length is directly related to the wavelength of radiation with the relation h≈λ/2. However, this formalism cannot be applied at optical frequencies because in this case the incident radiation is not entirely reflected from the metallic nanorods surface. The radiation penetrates into the metal and gives rise to free electron oscillations along this surface. Hence, nanoantennas operating at optical frequencies no longer directly respond to the external incident radiation's wavelength λ. Their dimensions depend on a shorter effective wavelength $\lambda_{eff} < \lambda$ that can be calculated by the nanoantenna's material properties. The simple linear scaling law for this effective wavelength had been derived in [15] and is given by:

$$\lambda_{eff} = n_1 + n_2 \left[\frac{\lambda}{\lambda_p}\right], \qquad (1)$$

where $\lambda_p$ is the plasma wavelength of the nanoantenna metallic material and $n_1$ and $n_2$ are two coefficients with length dimensions that depend on the antenna geometry and the material properties.

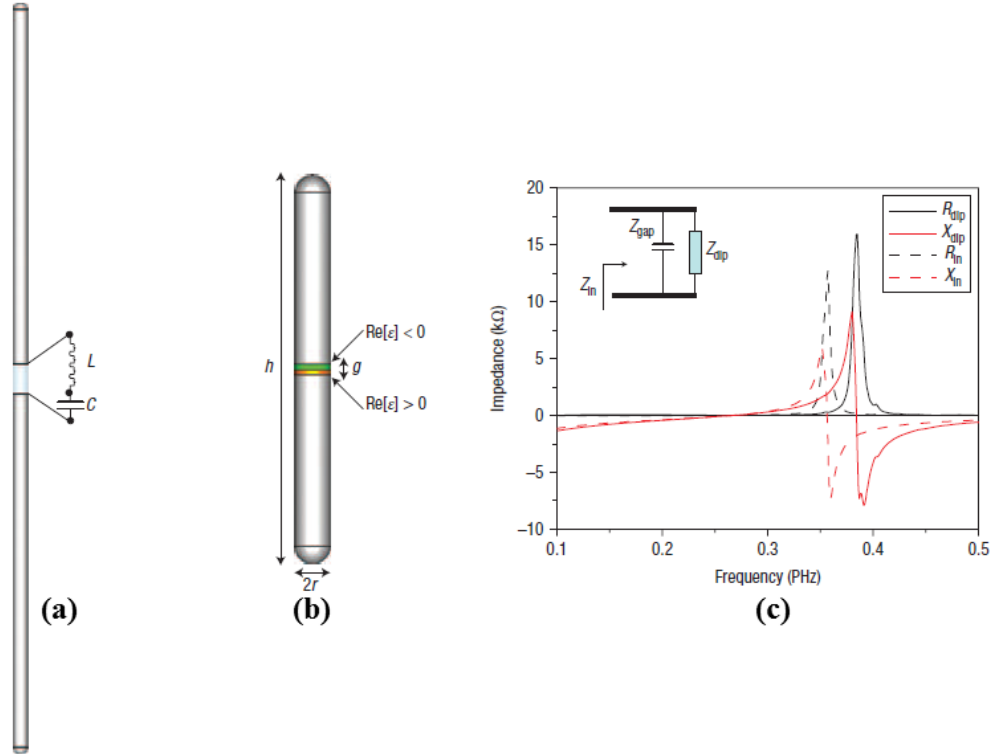

**Figure 4.** (a) RF dipole antenna loaded with lumped circuit elements. (b) Plasmonic nanodipole antenna loaded with inductive and capacitive nanoloads. (c) Input (dashed line) and dipole (solid line) complex impedances of the nanodipole shown in (b) [31].

Another interesting property of the RF dipole antenna is that its resonant frequency can be tuned by loading lumped circuit elements in the gap region, as it can be seen in Figure 4a. The translation of the tunable antenna concepts to the optical frequency regime is challenging but it can be done by incorporating lumped nanocircuit elements [17, 31]. The tuning of the dipole nanoantenna optical response can be achieved by adjusting the inductive and capacitive properties of its nanogap, similar to the tunable mechanisms used at RF dipole antennas [31]. Nanocircuit elements in the form of nanodisks can be used for this purpose, as it is shown in Figure 4b. They can be built by different materials like silver, gold or silicon. In general, metals (silver and gold) behave as nanoinductors and dielectrics (silicon) as nanocapacitors at optical frequencies [43]. The change in inductance and capacitance will lead to a change at the resonant frequency or, equivalently, wavelength of the optical nanoantenna. The radius of the nanodipole (r) is much smaller compared to the height (h) of the nanoantenna. The gap thickness is g and it is shown in Figure 4b. The impedance of a nanodisk with height t, diameter 2a and material permittivity ε, when excited by an electric field at frequency w, polarized parallel to its axis, is given by [31]:

$$Z_{disk} = \frac{it}{w\varepsilon\pi a^2}. \quad (2)$$

The input impedance of the nanoantenna system shown in Figure 4b is $Z_{in} = R_{in} - iX_{in}$, where $R_{in}$ and $X_{in}$ are the input resistance and reactance, respectively. The intrinsic dipole impedance is $Z_{dip} = R_{dip} - iX_{dip}$, where $R_{dip}$ and $X_{dip}$ are the dipole's resistance and reactance, respectively, when no material is placed at the nanogap region. The gap impedance is $Z_{gap}$ can be calculated using Eq. (2). Then, it can be properly de-embedded from the value of $Z_{in}$ to evaluate the intrinsic dipole impedance $Z_{dip}$. For a silver nanodipole with dimensions h = 110nm, r = 5nm and g = 3nm, $Z_{in}$ (dashed lines) and $Z_{dip}$ (solid lines) are computed and presented in Figure 4c. The intrinsic dipole impedance $Z_{dip}$ is found to be resonant around 356 THz. The zero value of $X_{dip}$ also arises at this resonant frequency point, which is near the height size condition h≈$\lambda_{eff}$/2 with $\lambda_{eff}$ < λ. This is consistent to the wavelength shortening effect in plasmonic nanoantennas, which was explained in previous paragraphs [15]. The effective wavelength is shortened to a much smaller value than the free space wavelength.

Materials with different permittivities can be placed at the nanogap and can be used as nanoloads to tune the resonance of the nanodipole antenna. Moreover, series and parallel combinations of these different nanoloads can be employed to further tailor and control the resonance of the nanodipole antenna. The resonant electric field response measured at the side of the nanodipole is shown in Figure 5a when the nanogap is filled with different nanoload materials, such as silver, gold, silicon, $SiO_2$ and $Si_3N_4$. Figure 5b presents the resonant electric field response of the nanodipole when the nanogap is filled with series and parallel nanoload combinations. Strong tunability in the resonant frequency of the nanodipole is obtained for all the loading cases shown in Figure 5. Two cylindrical nanodisk materials are assumed to be in series when they share a common interface, have the same radius $r_{in}$, and the excited electric field is normal to their common interface. Moreover, cylindrical nanoloads can be combined in a concentric way and create parallel loads in the nanogap. In the parallel configuration, the excited electric field is tangential to their common concentric interface. This is equivalent to the case of two parallel circuit elements, where both of them experience the same voltage drop at their edges. The impedance of the parallel combination of two concentric cylindrical nanorods with their common interface of radius r$_{in}$, total outer cell disk radius r, frequency w and gap g can be calculated by the parallel combination of the following impedances [31]:

$$Z_{inner\ load} = \frac{ig}{w\varepsilon\pi r r_{in}^2}, Z_{outer\ load} = \frac{ig}{w\varepsilon\pi(r^2 - r_{in}^2)}, \quad (3)$$

where Z$_{inner\ load}$ and Z$_{outer\ load}$ represent the load impedances of the inner and outer cell concentric nanodisks.

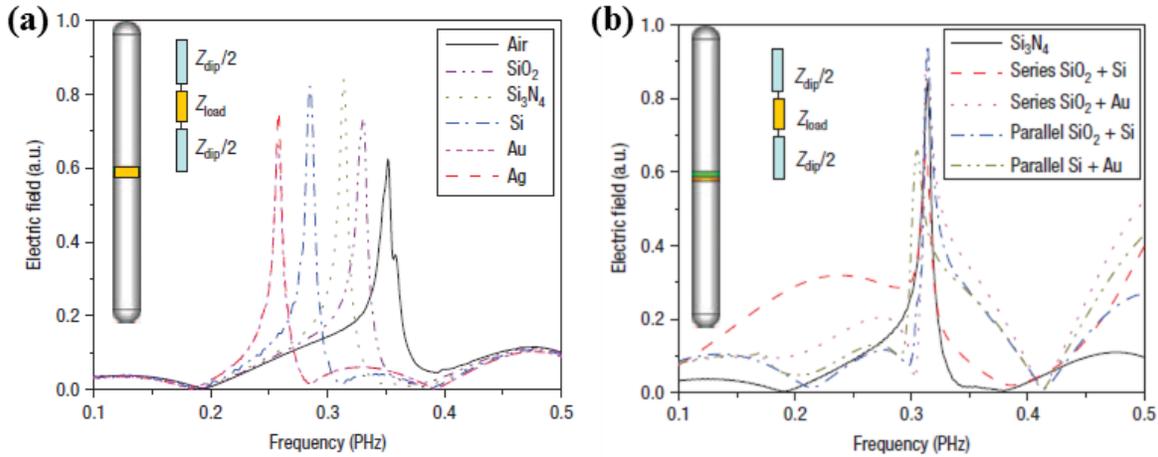

**Figure 5.** (a) Tunable electric field resonant response when the nanogap of the nanodipole shown in Figure 4b is loaded with different materials. (b) Tailoring the resonance of the nanodipole shown in Figure 4b using two nanoloads in parallel or series connection [31].

An alternative exciting application of linear nanoantennas is the potential to create ultrafast and broadband optical wireless communication channels. Nanoscale receiving and transmitting dipole antennas for wireless broadcast applications have been presented in [42]. The alternative configuration to these wireless channels is optical wired links which can be built by plasmonic metal-insulator-metal (MIM) waveguides. However, these MIM waveguides suffer from increased absorption along their metallic parts. Due to this severe limitation, they cannot provide long propagation distances at optical frequencies. To alleviate this problem, a nanoscale dipole nanoantenna wireless link shown in Figure 6a can be used instead of purely wired plasmonic waveguide channels.

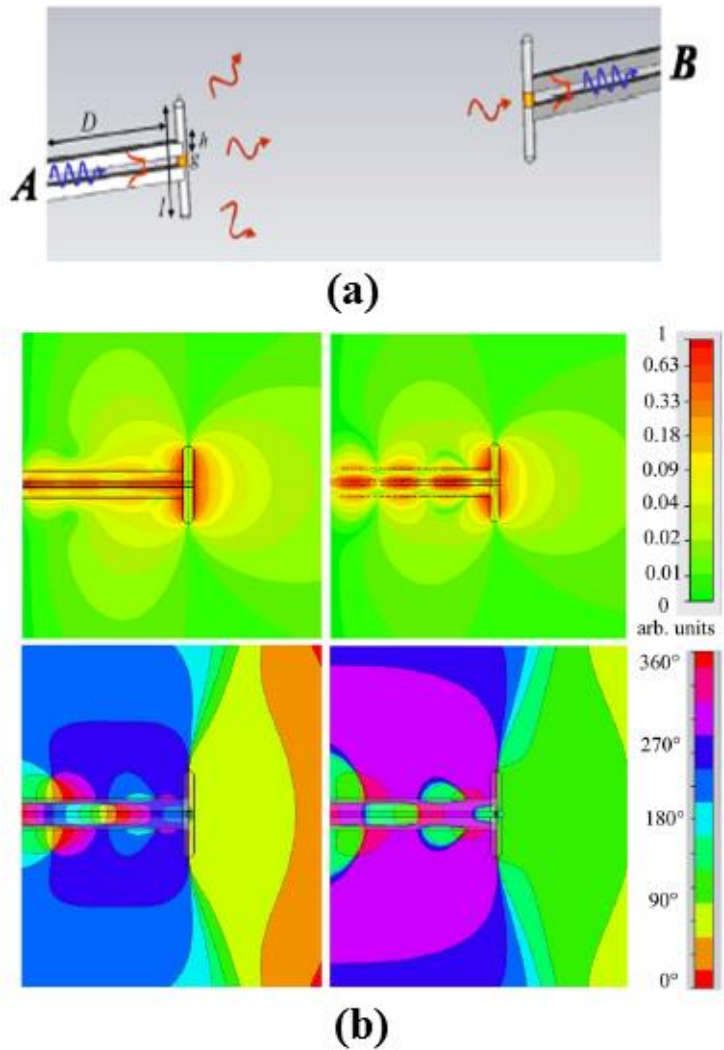

**Figure 6.** Wireless at the nanoscale: (a) Nanoscale wireless broadcast link with nanodipole antennas as transmitter and receiver. (b) Amplitude (top) and phase (bottom) of the magnetic field distribution for loaded (left) and unloaded (right) nanoantenna conditions [42].

Plasmonic waveguides act as feeds to the transmitting and receiving nanodipole antennas in the configuration shown in Figure 6a. The magnetic field distribution of the wireless link's transmitter is computed with full-wave electromagnetic simulations and is presented in Figure 6b for loaded (left) and unloaded (right) nanoantennas. The unloaded configuration corresponds to the case when the nanogap of the nanoantenna is filled with air. In the loaded configuration, the nanogap is filled with appropriate nanodisk loads in order the nanodipole's impedance to be tuned and matched to the surrounding free space. In the unloaded case, a substantial amount of the input power transmitted to the nanoantenna is reflected back to the feeding plasmonic waveguide. This is evident by the creation of strong standing waves at the feeding waveguide, as it can be seen in the right side of Figure 6b. This effect results in poor radiation performance. However, when the nanoantenna is loaded with appropriate nanodisks (loaded case shown in the left side in Figure 6b), excellent matching is achieved at the resonant frequency and the power transmitted from the plasmonic waveguide to the nanoantenna is efficiently radiated to free space with almost no back reflections.

## 3.2 Nonlinear nanoantennas

Until now, we have studied nanoantennas loaded with linear elements, such as nanoinductors and nanocapacitors, and their series and parallel combination. However, nanoantennas can also be loaded with nonlinear optical materials. In this configuration, boosted nonlinear functionalities can be achieved due to the strong confinement and enhancement of the far-field radiation at the nanoantenna's gap, where the nonlinear material will be placed. The design principles of nonlinear optical antennas are based on the same rich physics of the well-established research area of nonlinear optics. The work in this promising research field originated in the pioneering experiment of second harmonic generation observed by illuminating nonlinear crystals [44]. It exploits the nonlinear relationship between electric field E and resulting nonlinear polarization P [45-46]:

$$P = \varepsilon_0 [\chi^{(1)} E + \chi^{(2)} E^2 + \chi^{(3)} E^3 + \cdots]. \qquad (4)$$

In Eq. (4), $\varepsilon_0$ is the vacuum permittivity and $\chi(s)$ is the *s*th-order susceptibility of the nonlinear material. Numerous works by several researchers during the recent years have been dedicated to nonlinear nanoantennas, plasmonic metamaterials, and metasurface devices [47-68].

For example, the plasmonic nanopatch antenna can be used to enhance several nonlinear optical processes. It exhibits robust scattering response, which can be controlled by the material placed at the nanogap, combined with strong field enhancement at this nanoregion [28-30]. The design of a nonlinear plasmonic nanopatch antenna is presented in Figure 7a [66]. It is built by a silver nanocube with height l=80nm. The metallic nanocube is separated with the silver substrate by a thin spacer layer with thickness g=2nm. This spacer layer is made of Kerr nonlinear optical material $\chi^{(3)}$ with relative permittivity given by the relationship: $\varepsilon = \varepsilon_L + \chi^{(3)} |E|^2$, where $\varepsilon_L$=2.2, and $\chi^{(3)} = 4.4 \times 10^{-18}$ m$^2$ /V$^2$ [66]. This nonlinear material can be either semiconductor or polymer. The metallic silver substrate has height h=50nm and is placed above a glass semi-infinite slab with refractive index n=1.47. In this review paper, the nonlinear analysis of the film coupled nonlinear nanopatch antenna is limited to third order Kerr nonlinear effects in order to demonstrate optical hysteresis and all-optical switching processes. However, this structure can also have other nonlinear applications, such as enhanced second/third harmonic generation and four wave mixing. Note that the metallic parts of the plasmonic nanopatch antenna are simulated using the experimentally measured permittivity of silver [69].

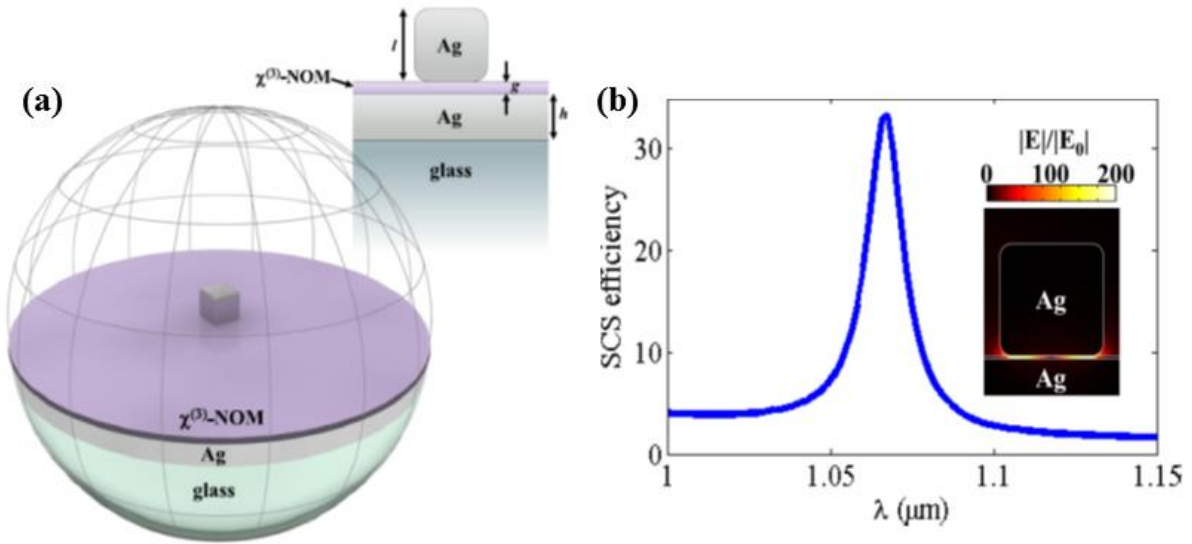

**Figure 7.** (a) Film coupled nonlinear nanopatch antenna. (b) SCS efficiency of the linear nanopatch antenna varying with the radiation wavelength. The amplitude of the field enhancement at the resonance is presented in the inset [66].

First, linear full-wave simulations of the nanopatch antenna system are performed after loading its thin spacer layer with a dielectric material having permittivity $\varepsilon_L$ = 2.2 and zero nonlinear part ($\chi^{(3)}$ =0). The dimensions of the nanoantenna are given in the previous paragraph. The scattering cross section (SCS) efficiency response at the linear operation is presented in Figure 7b. This result demonstrates a sharp narrowband scattering signature with a resonant scattering peak around 1.07 um. Standing waves are rapidly built inside the nanogap at this resonant frequency. The enhanced localized electric field amplitude distribution at the nanogap can be seen in the inset of Figure 7b. It has a maximum enhancement factor of 200.

Next, Kerr nonlinear $\chi^{(3)}$ material is loaded in the spacer layer to obtain bistable scattering response [66]. The intensity dependent permittivity of the Kerr nonlinear material is enhanced at the resonant frequency due to the ultrastrong fields confined at the nanogap. This enhancement will naturally lead to boosted nonlinear optical processes. Indeed, enhanced optical hysteresis and all-optical switching responses are obtained in Figures 8a and 8b, where the SCS efficiency as a function of wavelength and input pump intensity is plotted, respectively. The dashed lines in both plots show the unstable branch of the predicted nonlinear response. This unstable branch is the third solution of the nonlinear problem under study and it always leads to unstable, i.e. not physical, optical responses which can never be excited. In Figure 8a, the input pump intensity is kept fixed at 1MW/cm$^2$ and the SCS efficiency is plotted against the varying radiation's wavelength. Strong bistable performance is achieved due to the boosted fields at the nonlinear material placed in the nanogap. The nonlinear response can be further enhanced in case the input pump intensity is increased.

The SCS efficiency against the input pump intensity at a constant wavelength $\lambda=1.105\mu m$ is presented in Figure 8b. The proposed nonlinear optical nanoantenna design demonstrates strong all optical scattering switching behavior as the input power is varied. It is interesting that the presented results are obtained with low incident radiation power. The SCS efficiency remains constantly low until the input pump intensity reaches the threshold value of 820 kW/cm$^2$. The efficiency jumps up at this point and then it is kept constant as we further increase the input intensity. If we start decreasing the pump intensity below this threshold intensity, the nonlinear nanoantenna will keep its high SCS operation (see Figure 8b). When the input pump intensity is reduced below the lower threshold intensity of 180 kW/cm$^2$, the SCS will abruptly return to its initial low value state, similar to the scattering performance in the beginning of the illumination. This behavior can be utilized to build an efficient nanoswitch with the OFF mode placed at the low scattering branch and the ON mode positioned at the high scattering state.

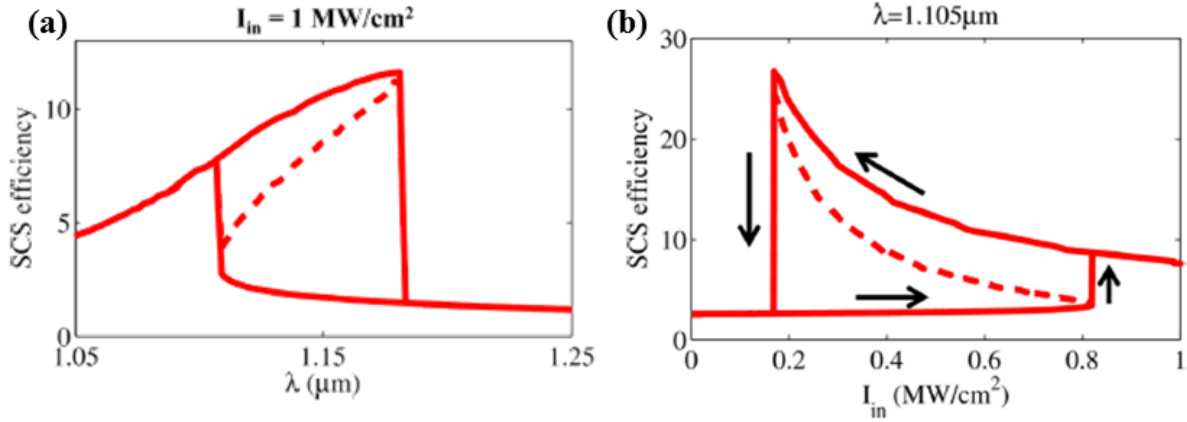

**Figure 8.** (a) SCS efficiency versus wavelength with the input intensity fixed at $I_{in} = 1 MW/cm^2$. (b) SCS efficiency versus input pump intensity for the wavelength fixed at $\lambda=1.105\mu m$ [66].

Optical bistability can also be obtained with other nonlinear nanoantenna configurations, such as nanodipole antenna arrays loaded with Kerr nonlinear materials at their nanogap [67]. In this set-up, the nonlinear performance can be further manipulated by the addition of a reflecting surface, which can be used to modify the coupling between the nanoantenna array and the incident light. Initially, we present the linear response of the nanoantenna array placed over a $SiO_2$ substrate. A single silver dipolar nanoantenna loaded with polystyrene and placed above the $SiO_2$ substrate is presented in Figure 9a. The nanoantenna arms are made of silver with $80\times20\times20 nm^3$ dimensions and are separated by 10nm gap. The gap is filled with polystyrene with refractive index $n_0=1.60$ and the nanoantenna is placed over a substrate of $SiO_2$ layer with refractive index $n=1.55$. The periodicity of the two dimensional array composed of these nanoantenna unit cells is 500nm in both in-plane directions. The total reflectance of this structure made of periodic silver nanodipoles is plotted as a function of wavelength in Figure 9b. The reflection response is calculated using the surface integral equation (SIE) technique [70, 71]. The SIE numerical method is particularly suited for the electromagnetic analysis of metallic and dielectric nanostructures with arbitrary shape [70, 71]. There is a strong increase in reflectance as the wavelength approaches 710nm, which corresponds to the resonant wavelength of the surface plasmon mode supported by the nanoantenna array.

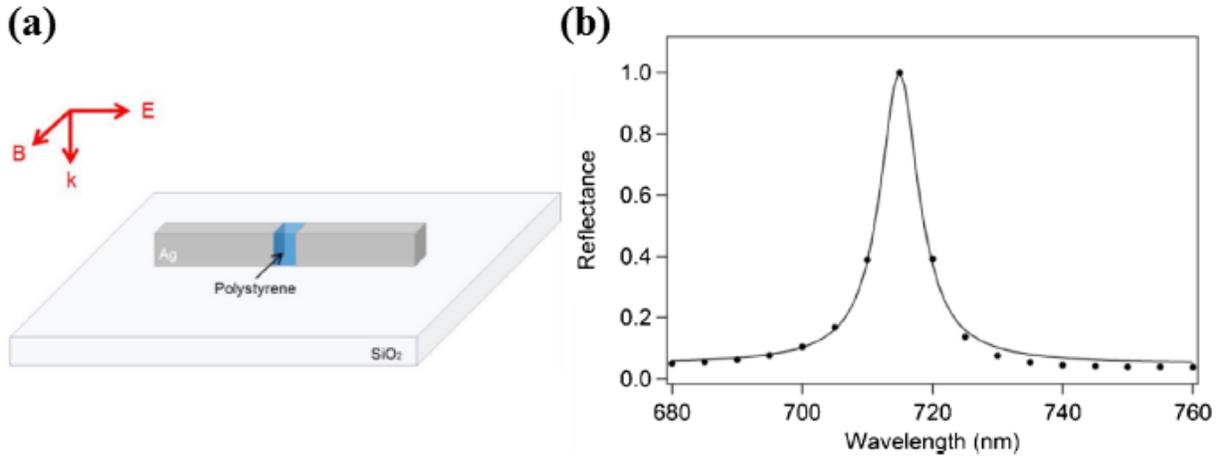

**Figure 9.** Linear response of the nanoantenna array. (a) Silver dipole nanoantenna loaded with polystyrene and placed over a SiO$_2$ layer substrate. This is the unit cell of the proposed periodic nanoantenna array. (b) Reflectance as a function of wavelength of the nanoantenna array [67].

The addition of a reflective surface below the nanoantenna array has a pronounced effect on its reflectance, which is now presented in Figure 10. In this case, a silver layer surface is added below the SiO$_2$ substrate layer of the nanoantenna unit cell shown before in Figure 9a. The unit cell of this new system is demonstrated in Figure 10a. The thickness of the SiO$_2$ substrate layer is denoted by d. The field intensity of the nanoantenna array unit cell at the incident resonant wavelength 710nm is also plotted in Figure 10a. The total reflectance of the nanoantenna array composed of the unit cell seen in Figure 10a is calculated as a function of the wavelength and is presented in Figure 10b for values of d ranging from 100nm to 200nm. The excitation of the surface plasmon resonance supported by the nanoantenna array leads to the formation of a pronounced dip at the reflectance spectra. Destructive interference of the incident and reflected wave is obtained at the nanoantenna position at the minimum reflection wavelength point. The coupling between the nanoantenna array and the incident electromagnetic wave is maximum at this frequency point and the surface plasmon resonance can be efficiently excited. The effect is analogous to Salisbury screens at microwave frequencies with the difference that the substrate's thickness is slightly shorter than $\lambda_{eff}/4$, since the incoming light penetrates into the silver reflector, which is not a perfect conductor at this wavelength, resulting in an additional phase shift.

For the largest thickness substrate (d=200nm), the reflectance does not depend on the wavelength anymore and remains constant and equal to one in the entire spectrum. This happens because the destructive interference and the field intensity at the nanoantenna's nanogap vanishes as the thickness of the substrate is increased. Hence, the surface plasmon resonance supported by this nanoantenna is not excited anymore and the total optical response is not modified, which leads to the flat reflectance spectrum for the case of 200nm thickness. The coupling between the nanoantenna array and the incident electromagnetic wave increases as we decrease the thickness d of the dielectric substrate.

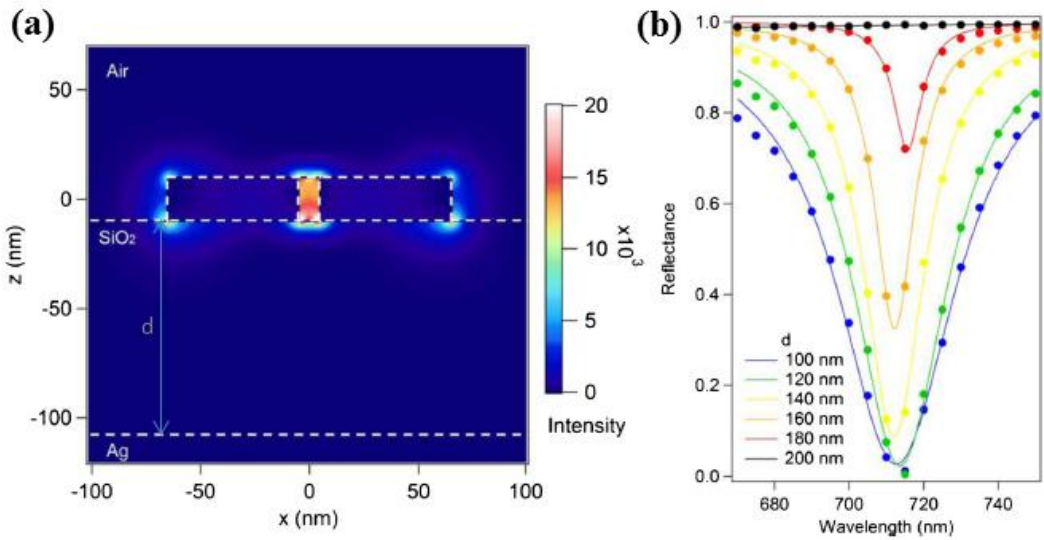

**Figure 10.** (a) Field intensity distribution of the nanoantenna unit cell placed over a SiO$_2$ substrate with thickness d and a silver reflecting surface. The incident resonant wavelength is λ=710nm. (b) Reflectance of the nanoantenna array as a function of the wavelength for different substrate thicknesses [67].

Subsequently, we study the nonlinear Kerr effect of the nanoantenna array. Polystyrene is loaded into the nanoantenna gap and its nonlinear response is triggered when the input intensity of the impinging radiation is increased. To accurately describe the Kerr nonlinearity of this material, the refractive index of polystyrene is given as a function of intensity [46]: $n = n_0 + n_2 I$, where $n_0$ is the linear part of the refractive index, $n_2$ is the nonlinear Kerr coefficient and I is the intensity inside the Kerr material loaded in the nanogap given by I=1/2ε$_0$|E|$^2$. The nonlinear Kerr coefficient is $n_2$=1.14×10$^{-12}$ cm$^2$/W [46], ε$_0$ is the permittivity of free space and E

is the electric field inside the nanogap. The nonlinear coefficient $n_2$ has positive value and the increase in the intensity of the input radiation will directly lead to an increase in the refractive index. Note that the nonlinear refractive index is an alternative expression of the nonlinear permittivity formula presented before. The dimensions of the nanoantenna array are the same with the linear case presented in the previous paragraph.

The nonlinear optical response resulting from loading the Kerr nonlinear material in the gap of the nanoantenna array is calculated using the analytical method presented in [68]. The nonlinear reflectance as a function of the incident intensity, considering the nonlinear Kerr effect described before, is plotted in Figure 11 at three wavelengths 710nm, 730nm and 770nm, and for two different $SiO_2$ thicknesses 100nm and 180nm. The first case shown in Figure 11a is for fixed input radiation wavelength at 710nm, corresponding to the surface plasmon resonance mode supported by the linear nanoantenna array, as it was shown before in Figure 10b. The reflectance increases as the incident radiation intensity increases for both considered $SiO_2$ thicknesses. However, optical bistability is not supported at this wavelength, which is clearly shown in Figure 11a. Next, we increase the wavelength to 730nm and observe the reflectance in Figure 11b as a function of the incident intensity. In this case and for 100nm thickness, again no optical bistability is observed, but for 180nm thickness strong optical bistability is observed in the reflectance spectrum. Finally, for 770nm wavelength input radiation, the reflectance is observed as a function of incident intensity in Figure 11c. Optical bistability is observed for both substrate thicknesses at this longer wavelength.

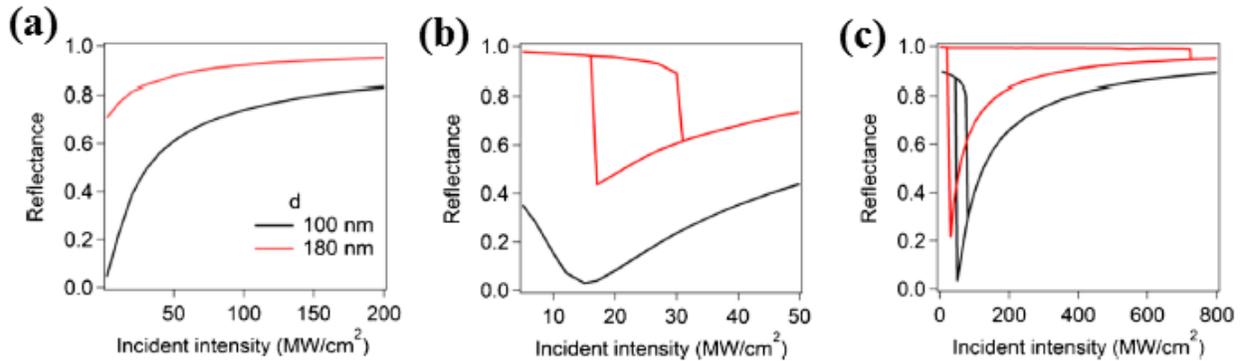

**Figure 11.** Nonlinear reflectance as a function of incident radiation intensity of a nonlinear nanodipole array with $SiO_2$ thicknesses 100nm (black line) and 180nm (red line) and at wavelengths: (a) 710nm, (b) 730nm, and (c) 770nm. Optical bistability is obtained due to the nonlinear Kerr effect of polystyrene which is loaded at the nanogap of each nanodipole [67].

### 3.3 Photodynamic applications of nanoantennas

Nanoantennas enable unprecedented control and manipulation of optical fields at nanoscale regions. Due to these properties, they can boost the performance of photo-detection, light emission and photoluminescence [72-87]. Moreover, they can also enhance the fluorescence and spontaneous emission rates of different emitters, which can lead to new nanophotonic applications [28, 74]. In this section, we will present several enhanced photodynamic processes based on nanoantenna configurations and their potential applications.

Emitters, like quantum dots and molecules, exhibit slow fluorescence and spontaneous emission rates. They cannot be used to create high speed optical communication components, such as ultrafast light-emitting diodes (LEDs) and other new optical sources. Recently, it was demonstrated that plasmonic nanopatch antennas are able to increase the inherent slow emission rates of these emitters [28, 74]. To enhance the spontaneous emission rates and to match their operation to high speed optical networks, the plasmonic nanopatch antenna system presented in Figure 12a was used [74]. In this configuration, the nanopatch antenna consists of a nanocube with 75nm size placed over a metallic thin film with thickness 50nm separated by a dielectric spacer layer of thickness 6nm where quantum dots are embedded, as shown in Figures 12a and 12b. The formation of an ensemble of similar single quantum dots placed at the gap of the nanopatch antenna is also presented in Figure 12c.

Modifying the photonic environment of these emitters with the use of plasmonic nanoantennas can lead to a rapid enhancement in their spontaneous emission rate, also known as Purcell factor. The enhancement in the local electric field at the nanopatch's gap can be directly translated to a boosted Purcell factor. The spontaneous emission rate enhancement and the radiative quantum efficiency of this plasmonic nanopatch antenna is calculated with full-wave simulations [74] and is shown in Figures 12d and 12e. Very high spontaneous emission rates are obtained, especially at the corners of the plasmonic nanocube used to form the nanoantenna system. Furthermore, the radiative quantum efficiency remains high (~50%) even with small plasmonic gaps (9nm in this case) due to the exceptionally high field enhancement at the nanogap. The quantum efficiency is calculated as the ratio of radiative over total spontaneous emission rate. It quantifies the fraction of the emitter's energy that is emitted as useful radiation and is not wasted in nonradiative (mainly thermal) processes. Thus, high radiative quantum efficiency is desirable for nanoantennas to be used as ultrafast and broadband optical communication components.

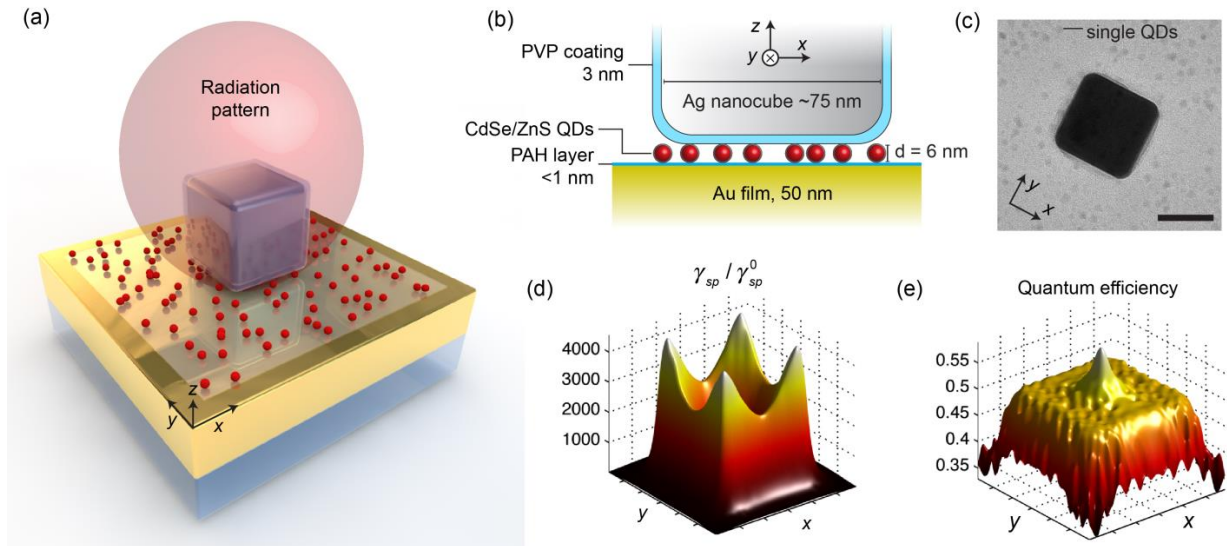

**Figure 12.** Nanopatch antenna with quantum dots. (a) 3D structure. (b) Front view. (c) Random ensemble of single quantum dots combined with the nanopatch antenna. (d) Spontaneous emission rate enhancement distribution. (e) Radiative quantum efficiency distribution [74].

The dark field scattering and fluorescence spectra of the nanopatch antenna system are demonstrated in Figures 13a and 13b, respectively [74]. In addition, the simulated and experimental scattering spectra of this plasmonic system are measured and shown in Figures 13c and 13d, when the gap of the nanopatch antenna is loaded with a polymer layer or quantum dots, respectively. The fundamental resonant mode exhibits a Lorentzian lineshape in the scattering response when a single nanopatch antenna without quantum dots in the nanogap region is used (see Figure 13c). On the other hand, the scattering spectrum broadens when quantum dots are embedded in the nanogap, as presented in Figure 13d [74]. The broadening in the scattering spectrum is caused due to the inhomogeneous dielectric environment in the nanogap region of the nanoantenna. This nonuniform dielectric environment is resulted due to the random spatial distribution of the embedded quantum dots. Finally, the narrow fluorescence spectrum of this nanoantenna system is also plotted in Figure 13d (red line).

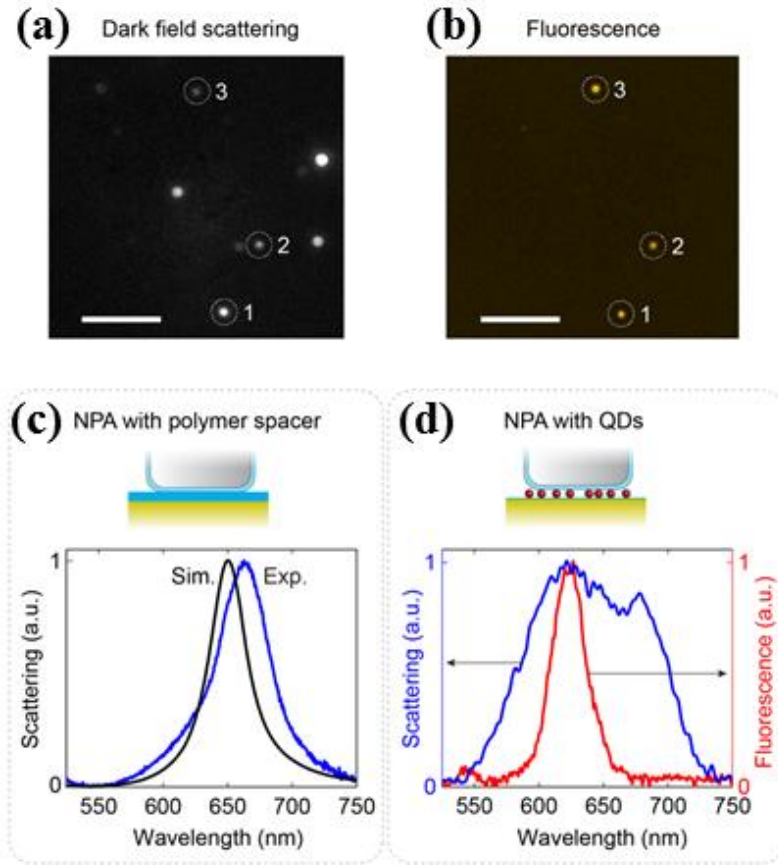

**Figure 13.** Spectral properties of the nanopatch antenna. (a) Scattering and (b) fluorescence images showing individual nanopatch antennas as bright spots with different intensities. (c) The scattering spectrum of the nanopatch antenna with a polymer spacer. (d) The scattering spectrum of the nanopatch antenna with quantum dots embedded in the spacer layer. The fluorescence spectrum is also shown in this plot (red line) [74].

The quantum dot fluorescence intensity is strongly enhanced by the coupling of quantum dots to plasmonic nanopatch antennas. It is presented in Figure 14a as a function of the average incident laser power for three different cases: quantum dots embedded in nanopatch antenna, quantum dots on glass substrate, and quantum dots on gold substrate. The intensity emitted from the quantum dots coupled to a single nanopatch antenna is substantially higher compared to the pure glass and gold substrate cases because strong fields are only induced with the nanoantenna configuration (Figure 14a). These strong fields are the main reason of the fluorescence enhancement in the nanoantenna/quantum dots plasmonic system. Note that quantum dots near the edges of the nanocube geometry experience higher field enhancement due to the shape of the resonant localized plasmon mode formed at the nanogap, as it was presented in previous sections.

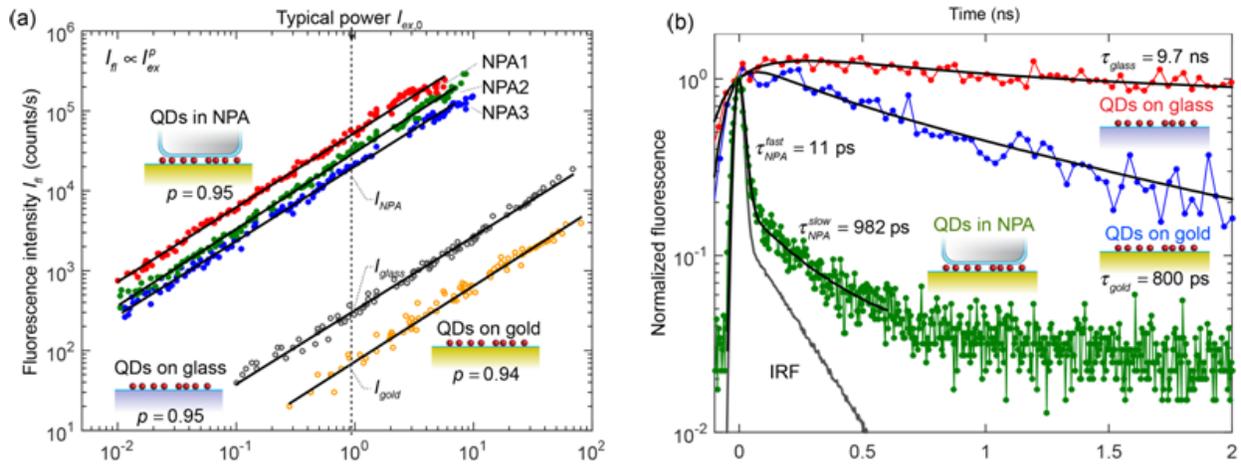

**Figure 14.** (a) Quantum dot fluorescence intensity as a function of the average incident laser power and (b) normalized time resolved fluorescence for three different configurations: quantum dots embedded in nanopatch antenna, quantum dots on gold substrate, and quantum dots on glass substrate [74].

Finally, the normalized time resolved fluorescence of the aforementioned structures (quantum dots embedded in nanopatch antenna, quantum dots on glass substrate, and quantum dots on gold substrate) is shown in Figure 14b. The enhancement in the spontaneous emission rate can be directly calculated by these time resolved fluorescence measurements. When quantum dots are coupled to the nanopatch antenna, the normalized fluorescence emission is drastically decreased in time compared to just placing quantum dots on glass or gold substrates. The substantial decrease in the fluorescence lifetime can be seen in Figure 14b. It is interesting that this rapid decrease in lifetime is accompanied by a simultaneous increase in the time-integrated fluorescence intensity shown in Figure 14a. These are ideal conditions to create ultrafast light sources for new optical communication networks.

Nanoantennas can also be used to enhance other reconfigurable photodynamic applications leading to the creation of efficient and ultracompact photoconductive optical switches, as it is presented in Figure 15 [75]. In this case, silver nanorods are used to build the suggested photoconductive nanodipole antenna (Figure 15). They have hemispherical end gap morphology. The nanodipole gap is loaded with amorphous silicon (a-Si) with a large electronic bandgap equal to 1.6eV. This large electronic bandgap is required in order to obtain an ultrafast tunable and reconfigurable response, which is important for optical switching applications. The substrate is not included in this geometry for simplicity.

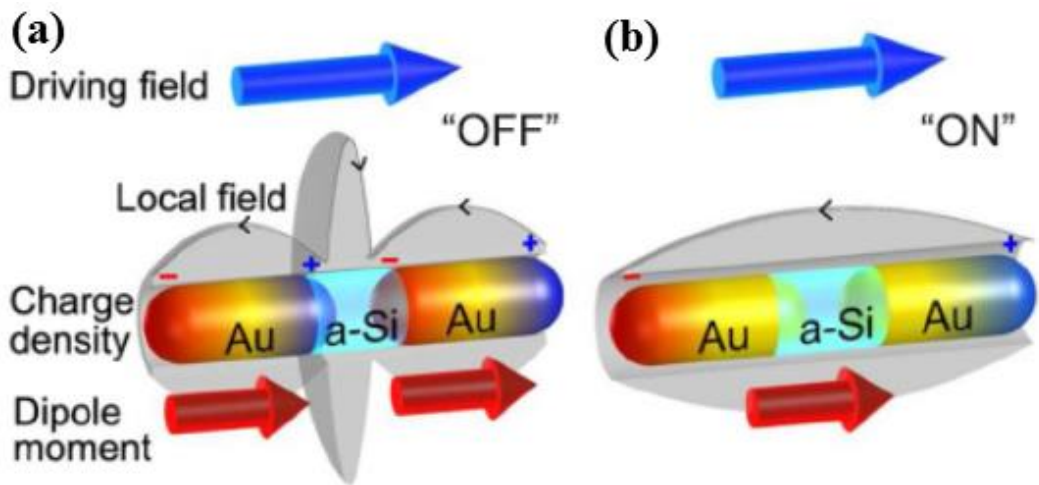

**Figure 15.** Reconfigurable switching operation of photoconductive nanoantenna. Operation at (a) switch "OFF" position and (b) switch "ON" position [75].

The a-Si material will have different properties when the doping level is increased. The tunable property of the material placed at the nanoantenna's gap will lead to a transition between capacitive and conductive operations. In the unswitched (OFF) case (low doping level of a-Si), the reconfigurable nanodipole antenna supports a half wavelength resonance over each individual arm and not over its entire geometry, as it is shown in Figure 15a. The same nanoantenna in the switched (ON) case (high doping level of a-Si) is shown in Figure 15b. Now, it operates above the free carrier switching threshold limit of amorphous silicon. In this case, the nanoantenna arms are conductively coupled and the nanodipole antenna supports a half wavelength resonance over the entire antenna length, similar to the usual dipole operation.

Nanostructured plasmonic nanoantennas can also be utilized for improved optical sensing and imaging applications, including surface-enhanced fluorescence and Raman scattering [79]. The surface enhanced fluorescence technique is based upon the design of particular engineered surfaces, such as nanoantenna arrays, in the vicinity of different emitters. These surfaces modify and control the local electromagnetic environment surrounding the emitter. This gives rise to an overall improvement in the fluorescence detection efficiency, similar to the nanopatch/quantum dot plasmonic system presented before. Raman scattering can also be boosted by nanoantennas [87]. This process is based on the inelastic scattering of photons, whose energy changes according to the vibration energy of a molecule. Moreover, optical nanoantennas can be used to enhance the efficiency of photovoltaic devices, i.e. for solar energy harvesting applications [80-81]. They have the potential to improve the absorption of solar radiation and lead to new plasmonic solar cell designs [88]. For example, an array of silicon or plasmonic nanowires can be patterned inside a thin film photovoltaic cell to boost the overall absorption of solar radiation while utilizing less than half of the required semiconductor material [81].

Another interesting application of optical nanoantennas will be in biological imaging. The schematic illustration of fluorescently labeled antibodies imaged by a nanoantenna probe is shown in Figures 16a and 16b [85]. The fluorescence patterns have different sizes and shapes, which originate from dissimilar antibody aggregates as they lie on the glass substrate. Lastly, optical antenna sensors can be helpful in spectroscopic applications [86]. For instance, the directivity enhanced Raman scattering using nanoantennas was demonstrated in [86].

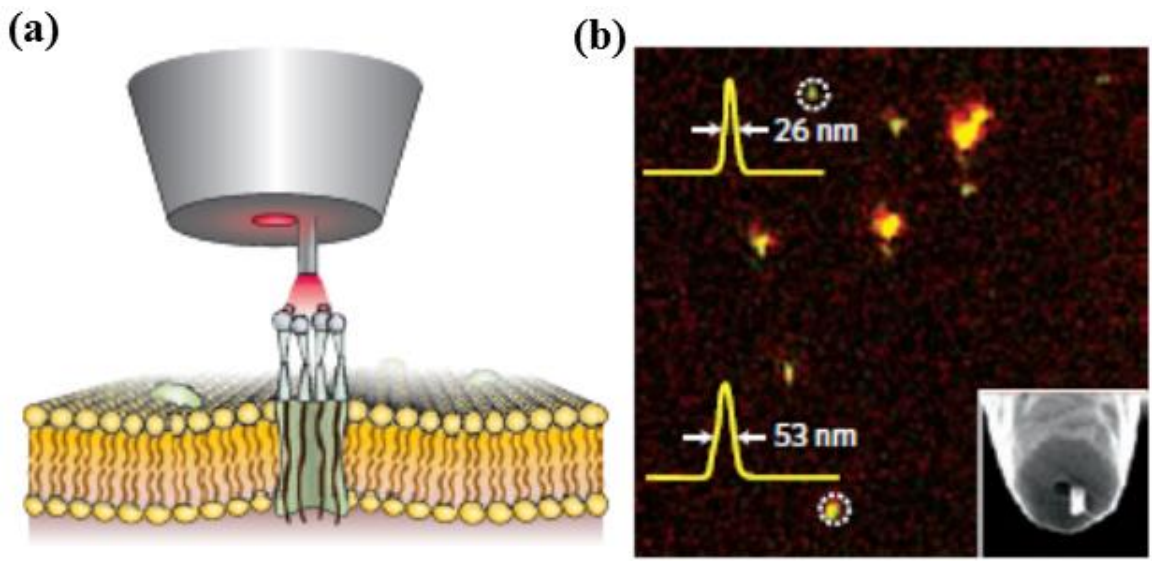

**Figure 16.** Nanoantenna applications in biological imaging. (a) Schematic illustration of imaging with probe based nanoantennas. (b) Florescence imaging with probe based nanoantennas. The geometry of the nanoantenna probe is shown in the inset [85].

**4 Conclusions**

We have reviewed several recent advances and applications of plasmonic nanoantennas. In particular, our review paper was focused on linear and nonlinear nanoantennas and their photodynamic applications. The ability to tailor the scattering response of nanoantennas using nanoloads has been reported. Furthermore, it was demonstrated that nonlinear plasmonic antennas are ideal candidates to build all-optical switching devices. Nanoantennas also hold promise for new optoelectronic and quantum information applications. In addition, photodynamic effects can be drastically enhanced with these plasmonic systems. Recently, the concept of plasmonic nanoantennas has also been extended to ultraviolet (UV) wavelengths with the use of aluminum nanorods [89] and plasmonic nanoparticles [90]. Finally, nanoantenna arrays can create compact metasurfaces [91-94], a new exciting research area which promises to revolutionize the practical applications of optical metamaterials. These new research fields are still in their infancy and innovative concepts and applications are expected to emerge in the near future.

*Acknowledgements.* This work was partially supported by the Office of Research and Economic Development at University of Nebraska-Lincoln.